# A New Method for Features Normalization in Motor Imagery Few-Shot Learning using Resting-State

M.Amin. Ghasemi[1], Sadjaad Ozgoli[2]*, Ali.M. NasrAbadi[3]

[1,2] Department of Electrical and Computer Engineering, Tarbiat Modares University, Tehran, Iran

[3] Department of Biomedical Engineering, University of Shahed, Tehran, Iran

**Abstract**

Brain-computer interface (BCI) systems are usually designed specifically for each subject based on motor imagery. Therefore, the usability of these networks has become a significant challenge. The network has to be designed separately for each user, which is time-consuming for the user. Therefore, this study proposes a method by which the calibration time is significantly reduced while the classification accuracy is increased. In this method, we calibrated the features extracted from the motor imagery task by dividing the features extracted from the resting-state into both open-eye and closed-eye modes and the state in which the subject moves his eyes. The best classification accuracy was obtained using the SVM classifier using the resting-state signal in the open eye, which increased by 3.64% to 74.04%. In this paper, we also investigated the effect of recording time of the resting-state signal and the impact of eye state on the classification accuracy.

**Keywords:** Few-Shot Learning, Subject-Independent, Motor Imagery (MI), electroencephalography (EEG), Brain-Computer Interface (BCI)

## 1. Introduction

Brain-Computer Interface (BCI) technologies and products establish the communication pathway between the human brain and the computer [1-2]. With the development of information technology and cognitive sciences, there has been a wave of attention and interest in converting mental activities into tangible signals and commands. The primary purpose of BCI studies is to develop a system that allows people with disabilities to interact with others and interact with the external environment and improve healthy people's abilities [3]. Technologies developing in this field seek to establish direct communication pathways between the brain and external devices, and their goal is to transfer human thoughts and intentions to the outside world. BCI products are often designed to support, augment, or enhance cognitive or sensorimotor processes.

---

[1] M.a.ghasemi@ieee.org
*Corresponding Author: [2] ozgoli@modares.ac.ir
[3] nasrabadi@shahed.ac.ir

Four types of signals are commonly used in BCI studies: slow cortical potentials (SCP) [4], steady-state visual evoked potentials (SSVEP) [5], p300 potentials in event-related potentials (ERP) [6], event-related desynchronization/event-related synchronization (ERD / ERS) based on motor imagery (MI) [7].

Brain-Computer Interface systems based on speech and motor imagery help people with speech and motor problems to control exoskeleton robots[8], move and guide wheelchairs [10-11], speak [12-13], and so on. BCI can be used as a tool to control the environment. This is possible by analyzing brain signals related to will and motor commands to execute a series of commands without muscles' need to intervene. In this way, BCI assistive robots can help disabled users in their daily lives and jobs [14].

Some of the early applications of BCIs target people with motor and speech impairments [15]. These applications help these people by providing alternative channels of communication. However, as this field expands, BCI is also entering the world of healthy people, acting as a physiological measurement tool that retrieves and uses emotional, cognitive, or feeling information from individuals [16].

Of course, MI-based BCI networks are not without problems. One of the problems with these networks is that the number of defined tasks is small [17]. These tasks include only the right-hand, the left hand, the tongue, and the foot imaging. The next problem is that BCI networks are designed and calibrated individually for each subject [18]. This means that we have to recalibrate the network for each new person. This can take hours for the user, and the reason for this problem is the non-stationary of the brain.

In this paper, we attempt to reduce system calibration time and provide a way to add to the processing steps by introducing a new approach to the classification of EEG signals based on user-independent motor imagery. This method involves the calibration of the feature vector using the feature vector's division over the feature vector extracted from the resting-state signal in three modes: open eye, closed eye, and moving eye. By providing this method, the problem of non-stationary EEG signals based on motor imagery is partially solved. Also, in this paper, the effect of eye condition and resting-state signal recording time on classifier performance has been studied. The rest of this article is organized as follows. Section (2) discusses the background of the work done in this area. Section (3) describes our proposed method. Section (4) describes the experimental paradigm and data recording. In Section (5), the results are reviewed, and it will be shown why our proposed method is successful. Section (6) discusses the strengths as well as the limitations that existed. Finally, in Section (7), the conclusions of this research are presented.

## 2. Related Work

This section explains the definitions of zero-shot learning, few-shot learning, and resting-state and discusses previous research on these topics.

## 2.1. Zero-shot learning

Zero-shot learning (ZSL) is one of the methods available in machine learning that allows the classifier to classify samples not present in the training phase. In many subjects, such as machine vision, natural language processing, and machine perception, much of the test phase data did not exist in the training phase, and the classifier has never seen this data before. Therefore, it is necessary to use methods such as ZSL to increase system accuracy [19].

S. Fazli et al. [20] Provided new BCI users with instant real-time control with little performance degradation. J. Cantillo-Negrete et al. [21] studied gender segregation in the design of user-independent networks. The results showed that gender separation has a positive effect on network performance. S. Hatamikia et al. [22] attempted to increase the user-independent network's performance by merely improving the algorithms. P. Gaur et al. [23] Proposed a method that uses the transfer learning model to solve the problem of poor classifier performance when there is non-stationary data. Their proposed method is a combination of the "multivariate empirical-mode decomposition" method, which is taken from 4 motor imagery tasks in the frequency band 8 to 30 Hz. A novel classification model that uses the structure of tangent space features from Riemannian geometry is. M.H. Mehdizavareh et al. [24] Sought to improve the BCI system's performance by providing a CCA-based approach to integrating subject-specific models and subject-independent information.

## 2.2. Few-shot learning

As the name implies, few-shot learning or one-shot learning refers to feeding a learning model with a tiny amount of training data instead of the regular practice of using a large amount of data.

S. An et al. [25], Considering the inter-subject variability and low SNR and the various challenges in extracting robust features in few-shot learning, proposed a method that could classify unseen subject categories even with limited MI EEG data. A.P. Costa et al. [26] Introduced an adaptive filter that combines the Common Spatial Patterns (CSP) filter and Recursive Least Squares (RSP) to update CSP filter coefficients. This adaptive algorithm is strengthened by the introduction of regularization using diagonal loading (DL). It will be able to reduce the training time of new data received from patients. S.M. Hosseini et al. [27] Presented a method based on an auto-adaptive calibration that uses the combination of EEG data space adaptation (DSA), a weighting scheme, a trial removal strategy to reduce the negative effects of non-stationary EEG signal. A. Singh et al. [28] Propose a new framework that transforms SPD matrices in lower dimensions through spatial filters regulated by EEG channels' prior information.

## 2.3. Resting-state

Resting-state EEG is a signal generated in the brain without performing any relief, and this signal may be visible at any frequency. A person's consciousness is detected by measuring and monitoring the resting-state EEG in many cases [29].

Y. Wang et al. [30] Proposed a zero-shot learning method that could be used to create a spatial filter for changes in the extracted EEG signal. They also obtained MI-based motor-based spatial filters by applying Independent Component Analysis (ICA) to motor imagery multi-channel EEG signal and resting-state. They then used these filters to classify the right and left hands. H. Zhang et al. [31] Have studied resting-state in motor imagery. Based on previous studies, they found the existence of Sensory and cognitive resting-state networks (RSNs), and they also came to two conclusions as they continued their research: sensory-motor and lateral visual networks exhibited more significant connectivity strengths in the precuneus and fusiform gyrus after learning; (2) Decreased network strength induced by learning was proved in the default mode network, a cognitive RSN. R. Zhang et al. [32] Investigated the relationship between Resting-state EEG and MI-BCI performance. In this study, they concluded that the network's spatial topologies and statistical measures are near related to the accuracy of motor imagery classification. They also concluded that node degree, mean functional connectivity, edge strengths, clustering coefficient, local efficiency, and global efficiency are directly related to classifier performance. At the same time, the characteristic path length is inversely related to classification accuracy. Y. Zheng et al. [33] Proposed a Resting-State Independent Component Analysis (RSICA)-based spatial filtering algorithm that aims to extract individual task-related spatial and temporal brain patterns from the resting-state data. M. Kwon et al. [34] proposed a modified predictor of MI-BCI performance that took into account both brain states (eyes- open and eyes- closed resting-states) and investigated this with 41 online MI-BCI session datasets acquired from 15 subjects.

## 3. Methodology

An overview of our proposed method can be seen in Fig. 1. In the training stage, after the preprocessing stage and the selection of the optimal channels, and the filtering of the data, feature extraction was performed. Then the same features were extracted from the resting-state.

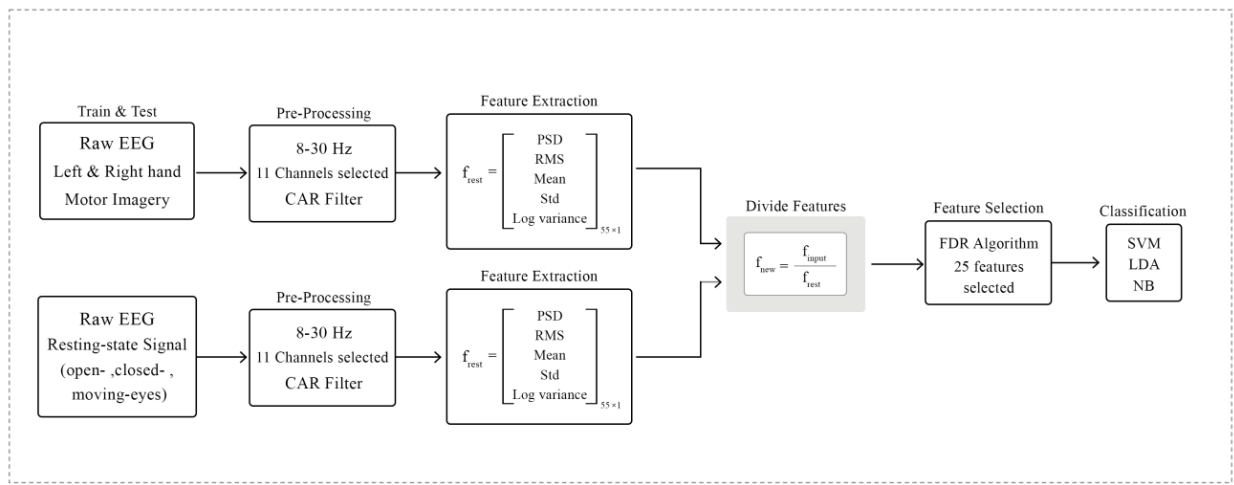

Fig. 1. The Framework of the Approach

After obtaining the feature vectors with the 2-class task and the resting-state features, these two vectors were divided. The result of this division is a calibrated feature vector. One feature vector was extracted separately for each subject, and these features are divided by the resting-state vector of the same subject. There is a calibrated feature vector for each subject that can be used for classification after the feature selection step.

In the testing stage, all the steps of the training phase are repeated. The difference is that the subject data selected as test data is not present in the training phase.

3.1. Feature Extraction

In this research, temporal features are extracted. From each channel, five features, including Power Spectral Density (PSD), Root Mean Square (RMS), mean, std, and log variance, were extracted [22-35-36].

3.2. Feature Selection

The FDR (Fisher's Linear Discriminant) algorithm is used in the feature selection step [37]. In this section, 25 features were selected from 55 features determined experimentally and by trial and error. The formula of this algorithm is shown in Eq. (1):

$$FDR = \sum_{i}^{M} \sum_{j \neq i}^{M} \frac{(\mu_i - \mu_j)^2}{\sigma_i^2 + \sigma_j^2} \tag{1}$$

where $i$ and $j$ are the characteristic mean and variance of the classes right and left hand.

3.3. Classification

After the feature extraction stage and the selection of 25 features from 55 features, the classification was carried out. At this time, we used the classifiers SVM, LDA, and NB [27-38]. This study aimed not to use improved classifications, and we did not look for a higher percentage of classifications. Of course, we could have used more classifiers, but we were content to choose more popular and simple classifiers.

## 4. Data Acquisition

4.1. Experimental paradigm

Dataset BCI Competition IV-2a is used in this article [39]. In this dataset, nine healthy persons are asked to sit in a comfortable chair and perform four imaging tasks, including left hand (class 1), right hand (class 2), feet (class 3), and tongue (class 4). Data recording was performed on two different days, and a total of 72 trials were recorded for each class. At the beginning of the recording, the participants recorded the EEG signal for 5 minutes while doing nothing. These 5 minutes are divided into three parts. 1) 2 minutes with eyes open 2) 2 minutes with eyes closed, and 3) 1 minute with eye movements. These 5 minutes are recorded for the effects of the EOG signal. Due to technical problems, no signal from subject 4 was recorded during these 5 minutes. This time is shown in Fig. 2.

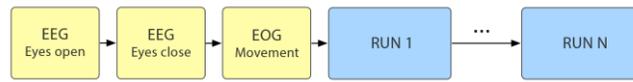

**Fig. 2.** Timing scheme of one session [39].

Participants were asked to sit in a comfortable chair in front of a monitor. At the beginning of a trial (t = 0s), a fixation cross appeared on the black screen. Besides, a short acoustic warning tone was presented. After two seconds (t = 2s), an arrow consisting of 4 modes: left, right, up, and down (the up arrow represented the tongue and the down arrow represented the feet) was displayed for 1.25s. After the sign was shown, participants were asked to imagine in their mind the task of the sign shown until the sign disappeared (t = 6s). After 6s, there was a short pause, and all these steps were repeated from the beginning. The paradigm is shown in Fig. 3.

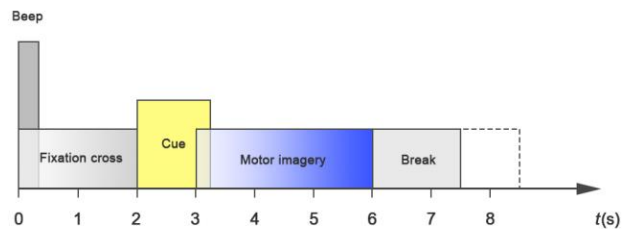

**Fig. 3.** Timing scheme of the paradigm [39].

4.2. Data recording

In this recording, twenty-two ag/agcl electrodes were used. The arrangement of these electrodes is shown in Fig. 4. The signals were sampled at 250 Hz and band-pass filtered between 0.5 Hz and 100 Hz. In addition to these 22 electrodes, three monopolar EOG electrodes with a sampling frequency of 250 Hz were used.

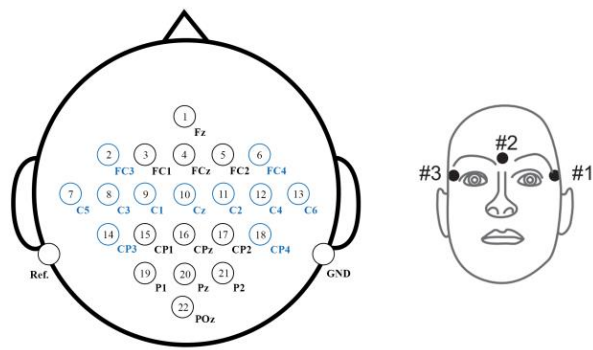

**Fig. 4** Left: Electrode montage corresponding to the international 10-20 system.
Right: Electrode montage of the three monopolar EOG channels [39].
**Note:** Blue circles are the selected channels.

## 5. Simulation and Result

In this section, we tested our proposed method using real EEG data. The obtained results have been examined in two ways: 1) The effect of eye condition on the resting-state signal 2) The effect of the recording time of the resting-state signal. Besides, we presented these results both visually and in tabular form.

5.1. Data Preprocessing

In this experiment, only two classes of the left and right hand were used. Since not all 22 EEG electrodes and even 3 EOG electrodes were needed, only the electrodes FC3, FC4, C5, C3, C1, Cz, C2, C4, C6, CP3, CP4 were used, as shown in Fig. 4. Then the Common Average Reference (CAR) filter was used [40]. An 8-30Hz 3-degree Butter-worth filter was used to filter the data [7].

5.2. Simulation

As mentioned, 72 trials are recorded per person in each class. Since this experiment aims to improve the subject's performance- independent motor imagery, all 144 trials per person (2×72=144) are used in training. In training, the data of 7 persons were used, which corresponds to 7×144=1008 trials. All data from one person, corresponding to 144 trials, were used for testing in each test. As mentioned earlier, subject four was excluded from the experiment because the resting-state data of user 4 was not recorded. The pseudo-simulation code is given in full in Table 2.

5.3. Resting-state

The resting-state EEG data were band-pass filtered from 8-30 Hz using FFT. This frequency band is the same band we used to filter the original EEG data. Then we used the CAR filter. Unlike many articles that only use the resting-state data of channels C3, C4, and Cz, we used the resting-

**Table 2**: Simulation Pseudo Code

- Load dataset: 2-class motor imagery and resting-state signal of each subject, (s1, s2, s3, s5, s6, s7, s8, s9)
- Apply CAR filter
- Apply band-pass filter: 8-30 Hz
- Channel Selection: selection 11 channels of 22 (FC3, FC4, C5, C3, C1, Cz, C2, C4, C6, CP3, CP4)
- Separate classes: right-hand and left-hand
- Separate train and test data: select 7 subjects as training data and 1 subject as test data
- Motor imagery tasks feature extraction: extract five features (PSD, RMS, mean, std, log var) from each channel, $F_{55 \times 1}^{MI\ task}$
- Resting-state feature extraction: extract five features (PSD, EMS, mean, std, log var) from each channel, $F_{55 \times 1}^{resting-state}$
- Divide task feature by resting-state features peer to peer, $F_{55 \times 1}^{calibrated}$
- Feature Selection: select 25 features out of 55, $F_{25 \times 1}^{calibrated}$
- Data normalization
- Train classifiers: SVM, LDA, NB
- Test stage

state data of all the channels we selected (FC3, FC4, C5, C3, C1, Cz, C2, C4, C6, CP3, CP4) [32-34].

After the preprocessing phase, we extracted from the EEG data the resting-state of each channel separately. Five features were mentioned in the methodology section. Eleven feature vectors were obtained for each subject. Then, we used these resting-state feature vectors to calibrate the task feature vectors.

5.4. Result

In this study, we examined the results of our proposed method from two aspects. First, we investigated the effect of using the resting-state signal in 4 modes using the SVM, LDA, and NB classifiers. 1) Classification without using the resting-state 2) Using the resting-state in open eyes mode 3) Using the resting-state with eyes closed 4) Using the resting-state with eye movement. These results can be seen in Table. 3. Second, we investigated the effect of the duration of recording the resting-state signal with eyes open only. This investigation was also performed in 4 cases. 1) Classification without using the resting-state 2) Using the recorded resting-state signal for 30 seconds 3) Using the recorded resting-state signal for 1 minute 4) Using the resting-state signal for 2 minutes. It should be noted that, as mentioned earlier, the resting-state signal is recorded once with the eyes open for 2 minutes. For our review, we separated the required amount from the same 2-minute signal at each step. These results can be seen in Table. 4. Of course, we must emphasize that the subject written in each column is the same one that was evaluated in the test phase, and the rest of the subjects were present in the training phase. So, we tested each network eight times with eight different subjects.

5.4.1. The result of using resting-state

Fig. 5(a) shows the classification results using the SVM classifier. As shown in Table 3, when comparing the mode of using the resting-state signal with the eye open to the mode where the resting-state signal was not used, we see an increase in classification accuracy for seven subjects, and only the accuracy of subject 8 was unchanged. The highest increase in classification accuracy was 4.87% for subject 2, and this increase was 11.11% for subject 5. The average classification accuracy increased by 3.64% and increased from 70.4% to 74.04%. Comparing the mode in which the resting-state signal was used with eyes closed to the mode in which the resting-state signal was not used, we see an increase in classification accuracy for all eight subjects. The highest increase in accuracy for subject 5 was 9.03%. The average classification accuracy also increased by 3.03% and reached 73.43% from 70.4%. Comparing the mode where the resting-state signal is used with the moving eye to the mode where the resting-state signal is not used, we see an increase in classification accuracy for all eight subjects. The highest increase in classification accuracy is seen in subject 5, with an increase of 6.94%. The average classification accuracy has also increased by 3.12%, reaching 73.52% from 70.4%.

Table 3: Classification results of all subjects in different eye modes in recording resting-state signal

| Subjects | Healthy subjects | | | | | | | | |
|---|---|---|---|---|---|---|---|---|---|
| | S1 | S2 | S3 | S5 | S6 | S7 | S8 | S9 | mean |
| SI SVM – No rest | 68.75 | 66.66 | 81.25 | 47.22 | 70.13 | 62.5 | 82.64 | 84.02 | 70.4 |
| SI SVM – rest (eye-open) | 71.53 | 71.53 | 84.03 | 58.33 | 72.92 | 65.28 | 84.72 | 84.02 | **74.04** |
| SI SVM – rest (eye-close) | 70.63 | 68.75 | 83.33 | 56.25 | 72.22 | 65.97 | 84.02 | 86.11 | 73.43 |
| SI SVM – rest (eye-movement) | 71.52 | 68.75 | 83.33 | 54.16 | 75 | 64.58 | 84.72 | 86.11 | 73.52 |
| SI LDA – No rest | 68.75 | 60.42 | 83.33 | 47.92 | 65.97 | 65.97 | 84.03 | 83.33 | 69.96 |
| SI LDA – rest (eye-open) | 67.36 | 70.14 | 82.64 | 56.25 | 70.83 | 63.19 | 82.64 | 84.03 | 72.13 |
| SI LDA – rest (eye-close) | 70.83 | 68.75 | 80.55 | 56.94 | 71.52 | 65.27 | 79.86 | 84.72 | 72.3 |
| SI LDA – rest (eye-movement) | 70.83 | 65.27 | 81.25 | 56.94 | 73.61 | 65.27 | 82.63 | 84.02 | **72.48** |
| SI NB – No rest | 61.80 | 64.58 | 78.47 | 56.25 | 72.22 | 61.11 | 77.08 | 84.72 | 69.53 |
| SI NB – rest (eye-open) | 64.58 | 65.97 | 81.25 | 56.94 | 72.92 | 61.11 | 84.72 | 83.33 | 71.35 |
| SI NB – rest (eye-close) | 63.88 | 68.05 | 77.77 | 56.25 | 74.3 | 61.11 | 84.72 | 85.41 | 71.44 |
| SI NB – rest (eye-movement) | 65.27 | 65.97 | 80.55 | 57.63 | 74.3 | 63.88 | 83.33 | 86.8 | **72.22** |

Fig. 5(b) shows the classification results using the LDA classifier. According to Table 3, compared with using the resting-state signal with eyes open with the mode of not using the resting-state signal, we see an increase in classification accuracy in 4 subjects and a slight decrease in 4 subjects. We observe the classification accuracy. The highest increase in classification accuracy was 9.72% in subject 2, 8.33% in subject 5, and 4.86% in subject 6. The average classification accuracy also increased by 2.17%, reaching 72.13% from 69.96%. Comparing the mode in which the resting-state signal was used with eyes closed to the mode in which the resting-state signal was not used, we see an increase in classification accuracy in 5 subjects. The highest increase in classification accuracy for subjects 2, 5, and 6 is 8.33%, 9.02%, and 5.55%, respectively. Comparing the mode of using the resting-state signal with the moving eye with the mode of not using the resting-state signal, we see an increase in classification accuracy for five subjects. The highest increase in classification accuracy for subjects 2, 5, and 6 is 4.85%, 9.02%, and 7.64%, respectively. The average classification accuracy has increased by 2.52% and has reached 72.48% from 69.96%.

Table 4: Classification results of all subjects at different times recording the resting-state signal.

| Subjects | Healthy subjects | | | | | | | | |
|---|---|---|---|---|---|---|---|---|---|
| | S1 | S2 | S3 | S5 | S6 | S7 | S8 | S9 | mean |
| SI SVM – No rest | 68.75 | 66.66 | 81.25 | 47.22 | 70.13 | 62.5 | 82.64 | 84.02 | 70.4 |
| SI SVM – rest (30s) | 68.75 | 70.83 | 83.33 | 54.86 | 75 | 65.97 | 81.94 | 87.5 | 73.52 |
| SI SVM – rest (1min) | 70.83 | 67.36 | 84.72 | 57.63 | 74.30 | 65.27 | 83.33 | 85.41 | 73.61 |
| SI SVM – rest (2min) | 71.53 | 71.53 | 84.03 | 58.33 | 72.92 | 65.28 | 84.72 | 84.02 | **74.04** |
| SI LDA – No rest | 68.75 | 60.42 | 83.33 | 47.92 | 65.97 | 65.97 | 84.03 | 83.33 | 69.96 |
| SI LDA – rest (30s) | 66.66 | 69.44 | 81.25 | 55.55 | 74.30 | 66.66 | 82.63 | 85.41 | 72.74 |
| SI LDA – rest (1min) | 66.66 | 68.05 | 81.25 | 54.16 | 73.61 | 65.97 | 81.94 | 81.94 | 71.70 |
| SI LDA – rest (2min) | 67.36 | 70.14 | 82.64 | 56.25 | 70.83 | 63.19 | 82.64 | 84.03 | 72.13 |
| SI NB – No rest | 61.80 | 64.58 | 78.47 | 56.25 | 72.22 | 61.11 | 77.08 | 84.72 | 69.53 |
| SI NB – rest (30s) | 63.19 | 65.97 | 81.94 | 58.33 | 70.83 | 63.19 | 85.41 | 84.72 | 71.70 |
| SI NB – rest (1min) | 63.88 | 66.66 | 80.55 | 58.33 | 73.61 | 61.80 | 84.02 | 84.72 | 71.70 |
| SI NB – rest (2min) | 64.58 | 65.97 | 81.25 | 56.94 | 72.92 | 61.11 | 84.72 | 83.33 | 71.35 |

Fig. 5(c) shows the classification results using the classifier NB. As shown in Table 3, compared to using the resting-state signal with the eye open and the mode in which the resting-state signal was not used, there is an increase in classification accuracy for six subjects. The highest increase in classification accuracy is seen in subject 8, with a value of 7.64%. The average classification accuracy has increased by 1.82% and has reached 71.35% from 69.53%. Comparing the mode in which the resting-state signal was used with eyes closed to the mode in which the resting-state signal was not used, we see an increase in classification accuracy in 5 subjects. The highest increase in accuracy for subject 8 was 7.64%. The average accuracy of classification has increased by 1.92% and has reached 71.44% from 69.53%. Comparing the mode of using the resting-state signal with the moving eye with the mode of not using the resting-state signal, we see an increase in classification accuracy in 8 subjects. The highest increase in accuracy for subject 8 is 6.25%. The average accuracy of classification has increased by 2.69% and has reached 72.22% from 69.53%.

As can be seen from the results, when using the SVM classifier, the best case is using the resting-state with the eye open, and when using the LDA and NB classifiers, the best case is using the resting-state with eye movement.

5.4.2. The effect of the recording time of the resting state signal

Fig. 6(a) shows the classification results of different resting-state time records using SVM. According to Table 4, as the resting-state signal recording time increased, the classification accuracy also increased in 3 subjects. Although there was a slight decrease in classification accuracy for the other subjects, the average classification accuracy increased with increasing signal recording time. The best signal recording time using the SVM classifier is 2 minutes

Fig. 6(b) shows the classification results of different resting-state recording times using the LDA. According to Table 4, increasing the resting-state signal's recording time did not necessarily improve classification accuracy. According to the results, the best signal recording time is the 30s. The average classification accuracy during this period is 72.74%.

Fig. 6(c) shows the classification results of different resting-state recording times with NB. According to Table 4, the classification accuracy decreases slightly as the resting-state signal recording time increases. This result shows that the signal recording time with this classifier is inversely related to the classification accuracy.

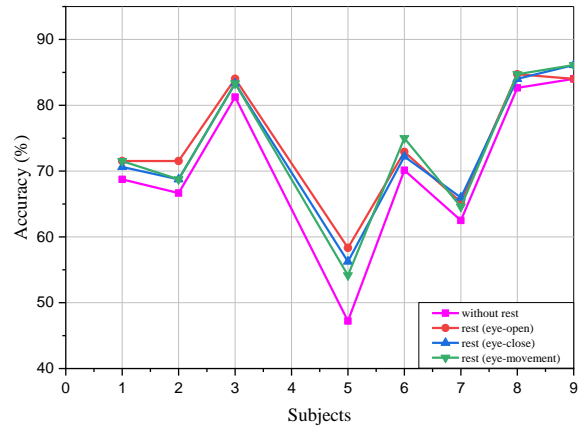

(a)

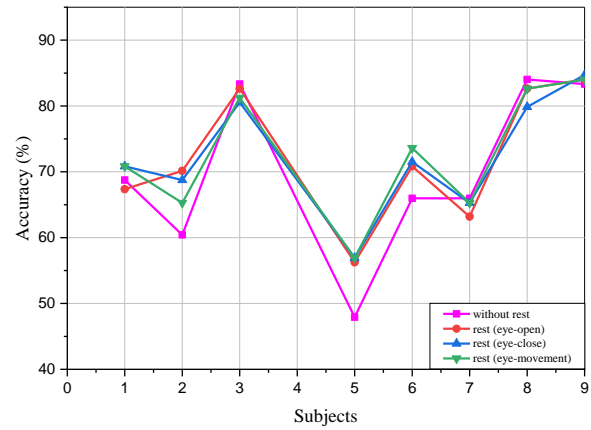

(b)

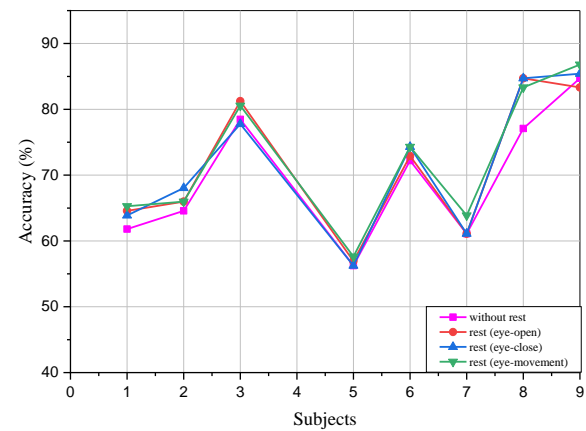

(c)

**Fig. 5** Classification results of all subjects in different eye modes in recording resting-state signal. (a) SVM classifier (b) LDA classifier (c) NB classifier

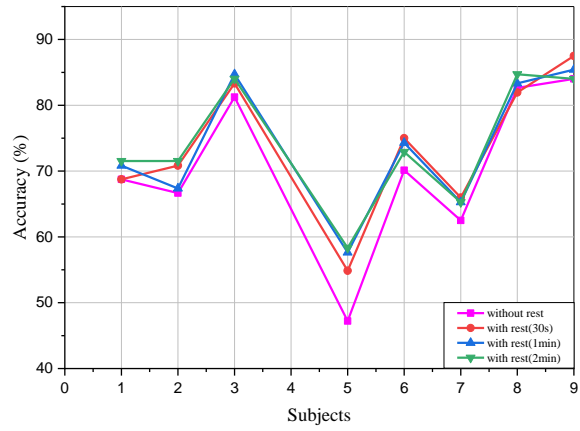

(a)

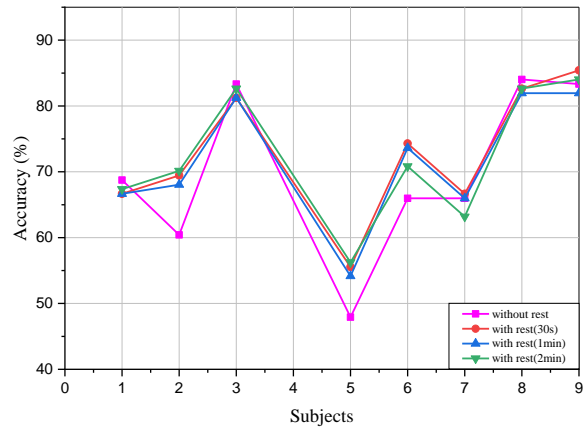

(b)

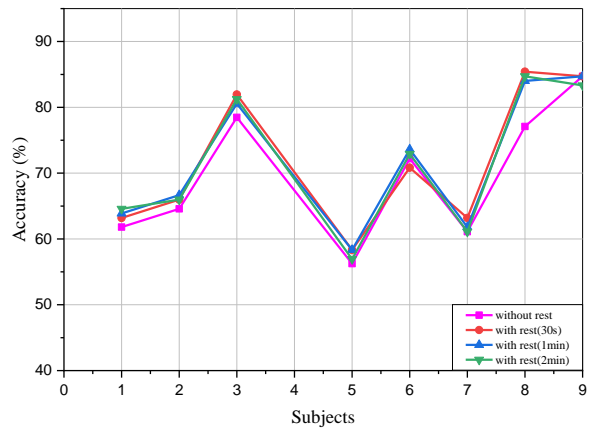

(c)

**Fig. 6** Classification results of all subjects at different times recording the resting-state signal.
(a) SVM classifier (b) LDA classifier (c) NB classifier

## 6. Discussion

The strength of this study was that we were able to provide a way to partially solve the problem of non-stationary EEG signals by recording the resting-state signal from the new subject in less than 2 minutes and improve the performance of the classifier without registering any other data from the new subject. However, some limitations could be addressed in the future:

1) In this study, the CSP filter was not used. This was because the CSP filter is usually designed before the feature extraction step. Therefore, it was not possible to design a CSP filter. This is because, in this case, the CSP filter was designed for uncalibrated features and was practically useless and reduced the classification accuracy. A way to design a CSP filter with calibrated features may be provided in the future.

2) In this work, an attempt was made to investigate the effect of using the resting-state signal, and for this reason, general features and classifiers were used. For this reason, improved methods can be used in future research to increase classification accuracy.

3) In this dataset, resting-state signals are recorded to investigate the EOG signal's effects on the EEG signal. For this reason, the recording of this signal may not have been done with the required accuracy. For a more detailed investigation and to complete the study on the effect of recording time of the resting-state signal, another dataset may be used in the future to repeat this investigation.

## 7. Conclusion

In this paper, we first attempted to implement the Few-Shot Learning method for the EEG-based motor imagery task. We did this in two ways: 1) First, we calibrated the features extracted from the motor imagery tasks by extracting the feature from the resting-state in the open-eye, closed-eyes, and eye movement. 2) Then, we investigated whether the resting-state's acquisition timing affects the classification accuracy. The results showed using the resting-state signal in each of the three modes, open-eye, closed-eye, and moving-eye improved classification accuracy. However, because three different classifiers were used in this study, the results were also different. When the SVM classifier was used, the highest classification accuracy was obtained in the open-eye mode. The best classification result was obtained with the resting-state signal in the moving eye mode using the LDA and NB classifiers. The best classification result using the resting-state signal was obtained using the SVM classifier and using the resting-state signal in the open eye mode, which increased the classification accuracy by 3.64% to 74.04%. We also investigated the effect of recording the relaxation time. Studies have shown that in this database, classification accuracy does not necessarily increase with increasing recording time.

We obtained from this experiment that when presenting a user-independent motion, it is possible to prepare the network for new subjects in less than 2 minutes by calculating the resting-state signal and performing the calibration, which positively affects the classification accuracy. Of course, like all normalization methods that have been proposed so far, this method may not always have a positive effect on all datasets. However, this method can be used as a new normalization method in user-independent motor imagery tasks and improve system performance.

8. **Conflict of Interest**

The authors have no conflict of interest relevant to this article.

[11] C. S. L. Tsui, J. Q. Gan, and O. Hu, "A self-paced motor imagery based brain-computer interface for robotic wheelchair control," Clin. EEG Neurosci., vol. 42, no. 4, pp. 225–229, 2011, doi: 10.1177/155005941104200407.

[12] K. Brigham and B. V. K. V. Kumar, "Imagined speech classification with EEG signals for silent communication: A preliminary investigation into synthetic telepathy," 2010 4th Int. Conf. Bioinforma. Biomed. Eng. iCBBE 2010, pp. 1–4, 2010, doi: 10.1109/ICBBE.2010.5515807.

[13] M. Matsumoto and J. Hori, "Classification of silent speech using support vector machine and relevance vector machine," Appl. Soft Comput. J., vol. 20, pp. 95–102, 2014, doi: 10.1016/j.asoc.2013.10.023.

[14] P. G. Vinoj, S. Jacob, V. G. Menon, S. Rajesh, and M. R. Khosravi, "Brain-controlled adaptive lower limb exoskeleton for rehabilitation of post-stroke paralyzed," IEEE Access, vol. 7, no. c, pp. 132628–132648, 2019, doi: 10.1109/ACCESS.2019.2921375.

[15] N. D. López, E. Monge Pereira, E. J. Centeno, and J. C. Miangolarra Page, "Motor imagery as a complementary technique for functional recovery after stroke: a systematic review," Top. Stroke Rehabil., vol. 26, no. 8, pp. 576–587, 2019, doi: 10.1080/10749357.2019.1640000.

[16] M. Bamdad, H. Zarshenas, and M. A. Auais, "Application of BCI systems in neurorehabilitation: A scoping review," Disabil. Rehabil. Assist. Technol., vol. 10, no. 5, pp. 355–364, 2015, doi: 10.3109/17483107.2014.961569.

[17] L. Duan et al., "Zero-shot learning for EEG classification in motor imagery-based BCI system," IEEE Trans. Neural Syst. Rehabil. Eng., vol. 28, no. 11, pp. 2411–2419, 2020, doi: 10.1109/TNSRE.2020.3027004.

[18] M. Arvaneh, I. Robertson, and T. E. Ward, "Subject-to-subject adaptation to reduce calibration time in motor imagery-based brain-computer interface," 2014 36th Annu. Int. Conf. IEEE Eng. Med. Biol. Soc. EMBC 2014, pp. 6501–6504, 2014, doi: 10.1109/EMBC.2014.6945117.

[19] Z. A. Yongqin Xian, Bernt Schiele, "Zero-Shot Learning - the Good, the Bad and the Ugly," IEEE Conf. Comput. Vis. Pattern Recognit., pp. 1–10, 2017.

[20] S. Fazli, F. Popescu, M. Danóczy, B. Blankertz, K. R. Müller, and C. Grozea, "Subject-independent mental state classification in single trials," Neural Networks, vol. 22, no. 9, pp. 1305–1312, 2009, doi: 10.1016/j.neunet.2009.06.003.

[21] J. Cantillo-Negrete, J. Gutierrez-Martinez, R. I. Carino-Escobar, P. Carrillo-Mora, and D. Elias-Vinas, "An approach to improve the performance of subject-independent BCIs-based on motor imagery allocating subjects by gender," BioMedical Engineering Online, vol. 13, no. 1. 2014, doi: 10.1186/1475-925X-13-158.

[22] S. Hatamikia, A. M. Nasrabadi, and N. Shourie, "Plausibility assessment of a subject independent mental task-based BCI using electroencephalogram signals," 2014 21st Iranian Conference on Biomedical Engineering, ICBME 2014. pp. 150–155, 2014, doi: 10.1109/ICBME.2014.7043911.